\documentstyle[preprint,aps,epsf]{revtex}
\newcommand {\bea}{\begin{eqnarray}}
\newcommand {\eea}{\end{eqnarray}}
\newcommand {\be}{\begin{equation}}
\newcommand {\ee}{\end{equation}}

\newcommand {\Dslash}{D\!\!\!/}

\begin{document}

%\draft
\preprint{SUNY-NTG-01-07, LBNL-47952}

\title{High Density Quark Matter under  Stress}

\author{P.~F.~Bedaque$^1$ and T.~Sch\"afer$^{2,3}$}

\address{$^1$ Nuclear Science Division, Lawrence Berkeley National 
Laboratory,\\ Berkeley, CA 94720\\
$^2$Department of Physics, SUNY Stony Brook,
Stony Brook, NY 11794\\ 
$^3$Riken-BNL Research Center, Brookhaven National 
Laboratory, Upton, NY 11973}

\maketitle

\begin{abstract}
  
 We study the effect of $SU(3)$ flavor breaking on
high density quark matter.
We discuss, in particular, the effect a non-zero electron 
chemical potential and a finite strange quark mass. We argue 
that these perturbations trigger pion or kaon 
condensation. The critical chemical potential 
behaves as $\mu_e\sim\sqrt{m m_s}\Delta/p_F$ and the critical 
strange quark mass as $m_s \sim m^{1/3}\Delta^{2/3}$, where 
$m$ is the light quark mass, $\Delta$ is the gap, and $p_F$ 
is the Fermi momentum. We note that parametrically, both the
critical $\mu_e$ and $m_s^2/(2p_F)$ are much smaller 
than the gap. 

\end{abstract}

\newpage

%%%%%%%%%%%%%%%%%%%%%%%%%%%%%%%%%%%%%%%%%%%%%%%%%%%%%%%%%%%%%%%%%%%%%%%%%
\section{Introduction}
\label{sec_intro}
%%%%%%%%%%%%%%%%%%%%%%%%%%%%%%%%%%%%%%%%%%%%%%%%%%%%%%%%%%%%%%%%%%%%%%%%%

 The study of hadronic matter in the regime of high baryon density 
and small temperature has revealed a rich and beautiful phase structure 
\cite{Rajagopal:2000wf,Alford:2001dt,Schafer:2000et}. One phase which
has attracted particular interest is the color-flavor locked (CFL) phase
of three flavor quark matter \cite{Alford:1999mk}. This phase is
expected to be the true ground state of ordinary matter at very
high density \cite{Schafer:1999ef,Schafer:1999fe,Evans:2000at}. 
State of the art calculations are not sufficiently 
accurate to predict the critical density of the transition to CFL
matter with any certainty. Current estimates typically range from 
$\rho_{crit}\sim (3-6)\rho_0$, where $\rho_0$ is the saturation
density of nuclear matter. An exciting prospect is the possibility
to put experimental constraints on the critical density from 
observations of neutron stars. Several proposals have been made
for observables that are characteristic of different superfluid
quark phases, and attempts are being made in order to include
these phases in realistic neutron star structure calculations
\cite{Page:2000wt,Blaschke:2000qx,Alford:2001sx}. 

 Initial work on the superfluid phases of QCD focussed mostly on 
idealized worlds with $N_f$ flavors of massless fermions and no 
external fields. But in order to understand the matter at the 
core of real neutron stars the effects of non-zero masses and
finite chemical potentials clearly have to be taken into account.
The first study of the effects of a non-zero strange quark mass on CFL 
quark matter was carried out in \cite{Alford:1999pa,Schafer:1999pb}.
The main observation in this work was that a finite strange quark
mass shifts the Fermi momentum of the strange quark with respect
to the Fermi momentum of the light quarks. If the mismatch between
the Fermi momenta is bigger than the gap then pairing between strange
and non-strange quarks is no longer possible. The transition from CFL 
matter to quark matter with separate pairing among light and strange 
quarks (2+1SC) is predicted to occur at $m_s\sim \sqrt{p_F\Delta}$. 
Alford et al.~observed that in the vicinity of this phase transition 
we expect to encounter inhomogeneous BCS phases \cite{Alford:2001ze} 
analogous to the Larkin-Ovchinnikov-Fulde-Ferell (LOFF) phase in 
condensed matter physics \cite{Larkin:1964,Fulde:1964,Abrikosov:1988}.
In the LOFF phase Cooper pairs have non-zero total momentum
and as a consequence, pairing is restricted to certain regions
of the Fermi surface. 

 In the present work we analyze CFL matter for strange quark 
masses and chemical potentials below the unlocking transition 
\cite{Rajagopal:2000ff}. We will argue that in this regime CFL 
matter responds to the external ``stress'' by forming a Bose 
condensate of kaons or pions \cite{Schafer:2000ew}. This effect 
can be understood as a chiral rotation of the CFL order parameter. 
Superfluid quark matter composed of only two flavors is 
characterized by an order parameter $\langle \epsilon^{abc}u^b 
C\gamma_5 d^c\rangle$ which is a flavor singlet 
\cite{Bailin:1984bm,Alford:1998zt,Rapp:1998zu}. This order parameter 
is ``rigid'' and superfluidity has to be destroyed in order to create 
a macroscopic occupation number of charged excitations 
\cite{Bedaque:1999nu}. CFL matter, on the other hand, is characterized 
by an order parameter which is a matrix in color and flavor space 
\cite{Alford:1999mk},
\be
\label{cfl}
\langle q_{L,i}^a C q_{L,j}^b\rangle
 = -\langle q_{R,i}^a C q_{R,j}^b\rangle
 = \phi \left(\delta_i^a\delta_j^b-\delta_i^b\delta_j^a\right),
\ee
where $i,j$ labels flavor and $a,b$ labels color indices.
We can introduce a chiral field $\Sigma$ that characterizes the 
relative flavor orientation of the left and right handed 
condensates \cite{Casalbuoni:1999wu}. In the vacuum $\Sigma=1$,
but under the influence of a perturbation $\Sigma$ may 
rotate. Because $\Sigma$ has the quantum numbers of 
pseudoscalar Goldstone bosons, such a rotation corresponds
to a macroscopic occupation number of Goldstone bosons.

\begin{figure}
\begin{center}
\leavevmode
\vspace{1cm}
\epsfxsize=10cm
\epsffile{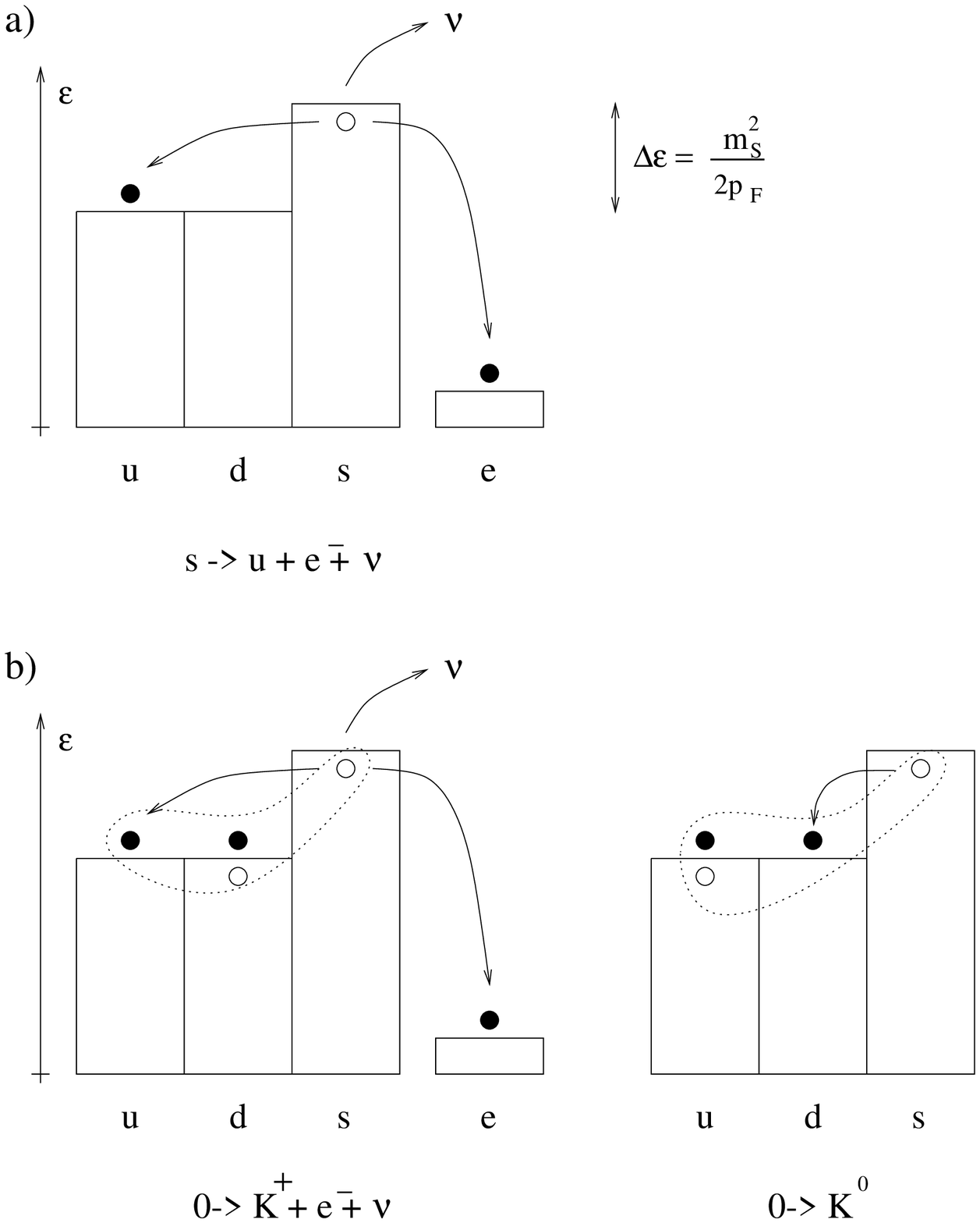}
\end{center}
\caption{\label{fig_dec}
Schematic picture of weak decays in normal (a) and superfluid (b) quark 
matter with three quark flavors. We assume that initially the density 
of all quark flavors is the same, so that $\epsilon_{F,s} \simeq 
\epsilon_{F,ud} + m_s^2/(2p_F)$. Solid and open circles show particles
$(p)$ and holes $(h)$. In (a) a strange particle decays into an up quark, 
an electron and a neutrino, leaving behind a strange hole. In the left 
panel of (b) a strange particle decays into an up quark, a down particle-hole 
pair, an electron and a neutrino. The remaining $(pp)(hh)$ configuration
has the quantum numbers of a $K^+$. In the right panel we show the 
decay of a strange quark into a $(pp)(hh)$ configuration with the 
quantum numbers of a $K^0$. }
\end{figure}

 There is an even simpler way to explain the phenomenon
of kaon condensation in superfluid quark matter, see
Fig. \ref{fig_dec}. Here we concentrate on the effect 
of a non-zero strange quark mass. A non-zero quark mass shifts
the energy of strange quarks in the vicinity of the Fermi
surface by $\sim m_s^2/(2p_F)$. In normal quark matter
this leads to the decay $s\to u+e^-+\bar{\nu}_e$ (or
$s\to u+d+\bar{u}$). This decay will reduce the number
of strange quarks and build up a Fermi sea of electrons
until the electron chemical potential reaches $\sim m_s^2
/(4p_F)$. In superfluid quark matter the system can also
gain energy $m_s^2/(2p_F)$ by introducing an extra up 
quark and a strange hole. This process appears to require
the breaking of a pair and therefore involve an energy cost 
which is of the order of the gap $\Delta$. This is not
correct, however. An up,down-particle/strange-hole pair has the 
quantum numbers of a kaon. This means that the energy cost is 
not $\Delta$, but $m_K\ll \Delta$. The CFL vacuum can decay 
into $K^+$ or $K^0$ collective modes via processes like 
$0\to (\overline{ds})(du)+e^-+\bar{\nu}_e$ or 
$0\to (\overline{us})(du)$.

 This paper is organized as follows. In section \ref{sec_3flavor}
we present general arguments for the existence of kaon and pion 
condensates in high density matter with broken flavor symmetry.
In section \ref{sec_match} we strengthen these arguments by 
performing an explicit matching calculation. In section \ref{sec_resp} 
we provide a different perspective on our results by using linear 
response theory.

%%%%%%%%%%%%%%%%%%%%%%%%%%%%%%%%%%%%%%%%%%%%%%%%%%%%%%%%%%%%%%%%%%%%%%%
\section{Three flavor quark matter at $m_s\neq 0$ and $\mu_e\neq 0$}
\label{sec_3flavor}
%%%%%%%%%%%%%%%%%%%%%%%%%%%%%%%%%%%%%%%%%%%%%%%%%%%%%%%%%%%%%%%%%%%%%%%

 In order to study QCD at high baryon density it is convenient to 
use an effective description that focuses on excitations close 
to the Fermi surface. Two effective descriptions of this type 
are available. The first effective theory is valid for excitation 
energies below the Fermi momentum $p_F$, while the second one applies 
to excitation energies below the gap $\Delta$. The coefficients
that appear in these effective theories can be worked out using 
matching arguments. In the first stage we match the microscopic 
theory, QCD at finite baryon density, to an effective theory below 
$p_F$. In the second step, we match this effective description to 
an effective theory involving Goldstone modes. 

 The QCD Lagrangian in the presence of a chemical potential
is given by
\be
\label{qcd}
 {\cal L} = \bar\psi \left( i\Dslash +\mu\gamma_0-\mu_e Q 
\gamma_0\right)\psi
 -\bar\psi_L M\psi_R - \bar\psi_R M^\dagger \psi_L 
 -\frac{1}{4}G^a_{\mu\nu}G^a_{\mu\nu},
\ee
where $M$ is a complex quark mass matrix which transforms as
$M\to LMR^\dagger$ under chiral transformations $(L,R)\in
SU(3)_L\times SU(3)_R$, $Q$ is the quark charge matrix,
$\mu$ is the baryon chemical potential and $\mu_e$ is (minus) 
the chemical potential for electric charge. As usual, we 
treat $M$ as a (spurion) field in order to determine the
structure of mass terms in the effective chiral theory.
Once this has been achieved, we set the mass matrix to 
its physical value $M={\rm diag}(m_u,m_d,m_s)$.

The quark field 
$\psi$ can be decomposed as $\psi=\psi_++\psi_-$ where 
$\psi_\pm=\frac{1}{2}(1\pm\vec{\alpha}\cdot\hat{p})\psi$. 
The $\psi_+$ component of the field describes quasi-particle 
excitations in the vicinity of the Fermi surface. Integrating 
out the $\psi_-$ field we get
\cite{Hong:2000tn,Hong:2000ru,Beane:2000ms}
\bea
\label{fs_eff}
S &=& \int \frac{dp_0}{(2\pi)}\frac{d^3p}{(2\pi)^3} \Bigg\{ 
 \psi_{L+}^\dagger \Big( p_0-\epsilon_p - v\cdot A \Big) \psi_{L+}
  - \frac{ \Delta}{2}\left(\psi_{L+}^{ai} C \psi_{L+}^{bj}
 \left(\delta_{ai}\delta_{bj}-
           \delta_{aj}\delta_{bi} \right) 
           + {\rm h.c.} \right) \nonumber \\ 
& & \hspace{2.0cm}\mbox{}
 +\psi_{L+}^\dagger \left(-\mu_e Q-  
        \frac{MM^\dagger}{2p_F}\right)  \psi_{L+}  
  +\frac{\bar\Delta}{8p_F^2} \psi_{L+}^{ai}C\psi_{L+}^{bj}
 \left(M^\dagger_{ai}M^\dagger_{bj}-
     M^\dagger_{aj}M^\dagger_{bi} \right)\nonumber \\ 
& & \hspace{2.0cm}\mbox{}
 + \left( R\rightarrow L, M\rightarrow M^\dagger, Q\rightarrow Q^\dagger
 \right)  + \ldots \Bigg\},
\eea
where $\epsilon_p=|\vec{p}|-\mu$, $v_\mu=(1,\vec{v})$ with $\vec{v}=
\vec{p}/p$, $\bar\Delta$ is a parameter that controls mass corrections
to the gap and $i,j,\ldots$ and $a,b,\ldots$ denote flavor and color indices.
In order to perform perturbative calculations in the superconducting 
phase we have added a tree level gap term $\psi_{L,R} C\Delta
\psi_{L,R}$ in the free part of the Lagrangian and subtracted it 
from the interacting part (not explicitly shown). The magnitude of 
$\Delta$ can be determined self consistently order by order in 
perturbation theory. In the normal phase both $\Delta$ and 
$\bar{\Delta}$ vanish. In this case, only the first mass term
in (\ref{fs_eff}) contributes. 

  We observe that at $O(1/p_F)$ flavor symmetry 
breaking due to a chemical potential for charge is indistinguishable 
from symmetry breaking due the quark mass matrix. Indeed, up to terms 
suppressed by additional powers of $(\Delta/p_F)$, $(p/p_F)$ or $(m/p_F)$
the Lagrangian (\ref{fs_eff}) is invariant under the time dependent 
flavor symmetry (from now on we drop the subscript ``+'')
\bea
\label{fs_eff_gauge}
\psi_{L} &\rightarrow&  L(t) \psi_{L}, \nonumber \\ 
\psi_{R} &\rightarrow&  R(t) \psi_{R}, \nonumber \\ 
\left(-\mu_e Q - \frac{M M^\dagger }{2p_F}\right)
  &\rightarrow& L(t) \left(-\mu_e Q - \frac{M M^\dagger }{2p_F} \right )
   L^\dagger(t) + i L(t) \partial_0 L^\dagger(t), \nonumber \\ 
\left(-\mu_e Q^\dagger - \frac{M^\dagger M}{2p_F}\right)
  &\rightarrow& R(t) \left(-\mu_e Q^\dagger -\frac{M^\dagger M}{2p_F} \right )
   R^\dagger(t) + i R(t) \partial_0 R^\dagger(t),
\eea
where $L(t)$ and $R(t)$ are left and right-handed time-dependent flavor 
transformations.

 For excitation energies below the gap $\Delta$ we can use an effective 
theory that includes only the pseudo-Goldstone bosons 
\cite{Casalbuoni:1999wu,Beane:2000ms,Son:1999cm,Manuel:2000wm}. 
The scale of the momentum and energy expansion in this theory
is set by the gap $\Delta$. Taking into account the symmetries
discussed above we see that a generic term in the  effective
lagrangian has the form
\be 
\label{chexp}
 {\cal L} \sim  f_\pi^2\Delta^2 \left(\frac{\partial_0-i\mu_e Q 
- iMM^\dagger/(2p_F)}{\Delta}\right)^n
\left(\frac{\vec\partial}{\Delta}\right)^m
\left(\frac{M M}{p_F^2}\right)^p
\left(\frac{\mu_e Q}{p_F}\right)^q.
\ee
This equation implies that the $N$'th order term in the effective 
lagrangian is given by the most general $SU(3)_L\times SU(3)_R$
invariant term constructed from the chiral field $\Sigma$ and 
containing $n$ covariant time derivatives, $m$ spatial derivatives,
$p$ powers of $M^2$, and $q$ powers of $\mu_e Q$ such that $N=
n+m+p+q$. We note that mass terms are suppressed by either 
$M^2/p_F^2$ or $MM^\dagger/(p_F\Delta)$. Terms of the form 
$M^2/p_F^2$ contain the quark mass matrix in the flavor 
anti-symmetric combination shown in the gap term in Eq.~(\ref{fs_eff}).

 The leading terms of the effective Lagrangian take the form
\bea
\label{leff2}
{\cal L}_{eff} &=& \frac{f_\pi^2}{4} {\rm Tr}\left[
 \nabla_0\Sigma\nabla_0\Sigma^\dagger - v_\pi^2
 \partial_i\Sigma\partial_i\Sigma^\dagger \right]
 + 2 A\left[\det(M) {\rm Tr}(M^{-1}\Sigma) + h.c.\right]+\ldots, \\
 & & \nabla_0\Sigma = \partial_0 \Sigma 
 + i \left(\mu_e Q +\frac{M M^\dagger}{2p_F}\right)\Sigma
 - i \Sigma\left(\mu_e Q^\dagger + \frac{ M^\dagger M}{2p_F}\right) .
\eea
Here $\Sigma=\exp(i\pi^a\lambda^a/f_\pi)$ is the flavor octet
chiral field and the $SU(3)_A$ generators are normalized as
${\rm Tr}[\lambda^a\lambda^b]=2\delta^{ab}$. We have not 
displayed the flavor singlet part of the effective lagrangian.
The first term in Eq.~(\ref{leff2}) is invariant under the 
approximate symmetry Eq.~(\ref{fs_eff_gauge}) because of the 
presence of the covariant time derivative. The second term
is not invariant under Eq.~(\ref{fs_eff_gauge}), but $A\sim 
f_\pi^2\Delta^2/p_F^2$ is suppressed by $1/p_F^2$, in 
accordance with Eq.~(\ref{chexp}).

 The $M^2$ term is not the most general term consistent with the 
symmetries. The structure of this term determined by the fact that 
it has to contain the quark mass matrix in a flavor anti-symmetric 
combination. $O(M^2)$ terms that are symmetric in flavor do not 
vanish, but they are strongly suppressed. We provide an estimate 
of these terms in App.~\ref{app_sym}.
 
 Despite the similarity between the effective theory for 
the Goldstone modes in the CFL phase and chiral perturbation 
theory in vacuum, there are important differences in the power 
counting. As usual, the contribution of loops is suppressed by 
powers of $p/f_{\pi}$. However, in the CFL phase $f_\pi\sim p_F
\gg \Delta$ which means that the suppression of loops with 
respect to tree level terms is much more pronounced than it
is in the vacuum. 

 More differences appear in the expansion in $M$. First of all, 
because of an approximate axial $Z_2$ symmetry in the CFL phase there 
are no odd powers in $M$. In addition to that, the $(MM^\dagger)
(M^\dagger M)$ terms can become comparable to the $M^2$ terms without 
breaking the chiral expansion. Indeed, as we shall argue below, this 
is likely to be the case for realistic values of $m_s$ and $p_F$. There 
are two reasons why the $(MM^\dagger)(M^\dagger M)$ term can 
become comparable to the $M^2$ term. First, the term proportional 
to $(MM^\dagger)(M^\dagger M)$ gives a contribution to meson masses 
which is of the order $m^2/p_F$ while the $M^2$ term contributes at 
order $m\Delta/p_F$. These contributions are comparable if $m\sim\Delta$, 
which is inside the regime of validity of the effective theory, 
$m < \sqrt{\Delta p_F}$. Second, in the realistic case where
$m_s \gg m_d,m_u$, the term quadratic in $M$ is proportional 
to at least one light quark mass, while the term quartic in $M$ 
contains terms proportional to $m_s^4$. 

 Using (\ref{leff2}) we can easily compute the masses of the Goldstone 
bosons in the CFL phase. At large density Lorentz invariance is broken
and we identify the mass with the energy of a $\vec{p}=0$ mode. For
$\mu_e=0$ the masses of the flavored states are given by
\bea 
\label{mgb2}
 m_{\pi^\pm} &=&  \mp\frac{m_d^2-m_u^2}{2p_F} +
         \left[\frac{4A}{f_\pi^2}(m_u+m_d)m_s\right]^{1/2},\nonumber \\
 m_{K_\pm}   &=&  \mp \frac{m_s^2-m_u^2}{2p_F} + 
         \left[\frac{4A}{f_\pi^2}m_d (m_u+m_s)\right]^{1/2}, \\
 m_{K^0,\bar{K}^0} &=&  \mp \frac{m_s^2-m_d^2}{2p_F} + 
         \left[\frac{4A}{f_\pi^2}m_u (m_d+m_s)\right]^{1/2}.\nonumber
\eea
The splitting between particles and anti-particles can be understood by 
observing that the crossed terms in the kinetic term of Eq.~(\ref{fs_eff}) 
act as an effective chemical potential for strangeness/isospin even 
if $\mu_e=0$. We observe that the pion masses are not strongly affected 
but the mass of the $K^+$ and $K^0$ is substantially lowered while
the $K^-$ and $\bar{K}^0$ are pushed up. As a result the $K^+$
and $K^0$ meson become massless if $m_s\sim m_{u,d}^{1/3}\Delta^{2/3}$.
For larger values of $m_s$ the kaon modes are unstable, signaling the 
formation of a kaon condensate. 

 Once kaon condensation occurs the ground state is reorganized.
For simplicity, we consider the case of exact isospin symmetry
$m_u=m_d\equiv m$. The most general ansatz for a kaon condensed 
ground state is given by
\bea
\label{k0+_cond}
\Sigma &=& \exp\left(i\alpha \left[
  \cos(\theta_1)\lambda_4+\sin(\theta_1)\cos(\theta_2)\lambda_5
  \right.\right.\nonumber \\
 & & \hspace{1.5cm}\left.\left.\mbox{}
   + \sin(\theta_1)\sin(\theta_2)\cos(\phi)\lambda_6
   + \sin(\theta_1)\sin(\theta_2)\sin(\phi)\lambda_7
   \right]\right).
\eea
With this ansatz the vacuum energy is given by
\be 
\label{k0+_V}
 V(\alpha) = -f_\pi^2 \left( \frac{1}{2}\left(\frac{m_s^2-m^2}{2p_F}
   \right)^2\sin(\alpha)^2 + (m_{K}^0)^2(\cos(\alpha)-1)
   \right),
\ee
where $(m_K^0)^2= (4A/f_\pi^2)m_{u,d} (m_{u,d}+m_s)$ is the $O(M^2)$ 
kaon mass in the limit of exact isospin symmetry. Minimizing the vacuum 
energy we obtain $\alpha=0$ if $m_s^2/(2p_F)<m_K^0$ and $\cos(\alpha)
=(m_K^0)^2/\mu_{eff}^2$ with $\mu_{eff}=m_s^2/(2p_F)$ if $\mu_{eff}
>m_K^0$. We observe that the vacuum energy is independent of 
$\theta_1,\theta_2,\phi$ even if $\alpha\neq 0$. This implies
that the effective potential in the kaon condensed phase has
three flat directions. The hypercharge density is given by
\be 
n_Y = f_\pi^2 \mu_{eff} \left( 1 -\frac{m_K^4}{\mu_{eff}^4}\right),
\ee
where $\mu_{eff}=m_s^2/(2p_F)$. This result is typical of a 
weakly coupled Bose gas \cite{Son:2000xc,Kogut:2000ek,Kogut:2001id}.
We also note that within the range of validity of the effective theory, 
$\mu_{eff}<\Delta$, the hypercharge density satisfies $n_Y<\Delta 
p_F^2/(2\pi)$. The upper bound on the hypercharge density in the
condensate is equal to the particle density contained within a strip 
of width $\Delta$ around the Fermi surface. 

 The symmetry breaking pattern is $SU(2)_I\times U(1)_Y\to U(1)$ 
where $I$ is isospin and $Y$ is hypercharge. It is amusing to note 
that this is the symmetry breaking pattern of the standard model. 
Kaon condensation is analogous to electroweak symmetry breaking
with a composite Higgs field \cite{Farhi:1981xs,Kaplan:1984fs}.
We can discuss kaon condensation in terms of an effective field
theory which only involves a complex kaon doublet $\Phi = (K^+,K^0)$
\be 
\label{higgs}
{\cal L} = 
   \big[\left(\partial_0+i\mu_{eff}\right)\Phi^\dagger\big] 
    \big[\left(\partial_0-i\mu_{eff}\right)\Phi\big]
   -(m_{K}^0)^2\left(\Phi^\dagger\Phi\right) 
   -\lambda \left(\Phi^\dagger\Phi\right)^2 .
\ee
If $\mu_{eff}>m_{K}^0$ the kaon
field acquires a non-zero vacuum expectation value $\langle\Phi\rangle 
= (0,v)$ and the $SU(2)\times U(1)$ symmetry is broken to $U(1)$. 
From (\ref{higgs}) we get $v=(\mu_{eff}^2-(m_K^0)^2)/(2\lambda)$. We 
can fix $\lambda$ by comparing the amplitude of the kaon field
to the result obtained from the chiral theory. We find $\lambda=
(m_K^0)^2/(2f_\pi^2)$. 

\begin{figure}
\begin{center}
\leavevmode
\vspace{1cm}
\epsfxsize=10cm
\epsffile{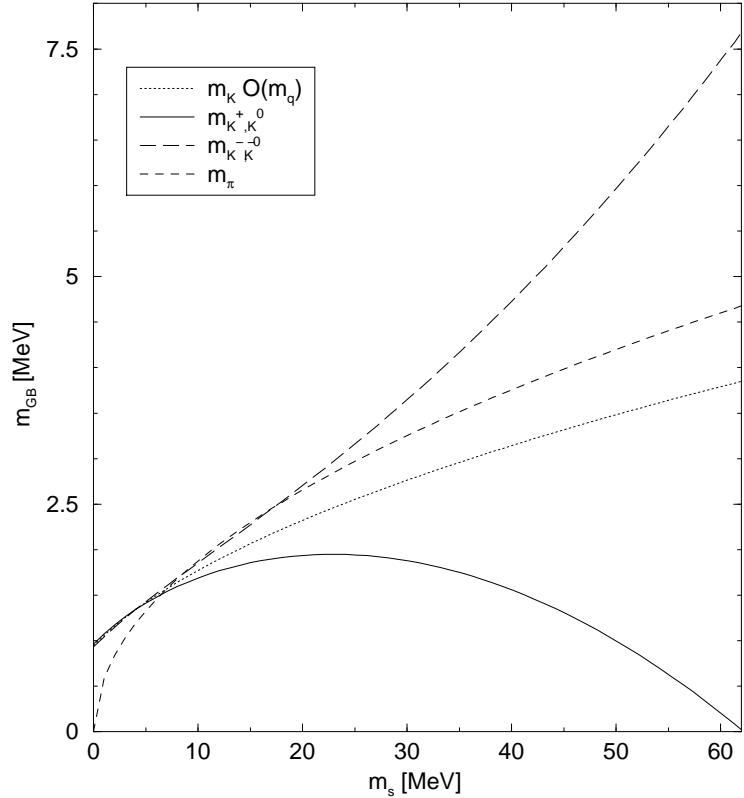}
\end{center}
\caption{\label{fig_kaon_ms}
Masses of $K^\pm$ and $K^0,\bar{K}^0$ excitations in the color-flavor 
locked phase. We show the excitation energies as a function of $m_s$
for $p_F=500$ MeV. The gap $\Delta=67$ MeV and the pion decay constant
$f_\pi=104$ MeV were determined to leading order in perturbation theory.
The solid and dashed curve show the masses of the $(K^+,K^0)$ and 
$(K^-,\bar{K}^0)$ states. The dotted curve shows the kaon masses  
calculated from the leading order $O(m_q)$ term. The short dashed
curve shows the pion masses.}
\end{figure}

 In weak coupling the coefficients of the effective Lagrangian can be 
computed and more quantitative statements about the onset of kaon 
condensation can be made. The gap is given by 
\cite{Schafer:1999jg,Pisarski:2000tv,Hong:2000fh,Brown:1999aq,Schafer:1999fe}
\be 
\label{gap}
 \Delta = 512\pi^4 2^{-1/3}(2/3)^{-5/2}\mu g^{-5}
 \exp\left(\frac{3\pi^2}{\sqrt{2}g}\right).
\ee  
The pion decay constant $f_\pi$ has been 
computed to leading order in $\alpha_s$ \cite{Son:1999cm} (a factor  
$2$ discrepancy in the literature will be resolved in section 
\ref{sec_match})
\be
 f_\pi^2=\frac{21-8\log 2}{18}\frac{\mu^2}{2\pi^2}.
\ee
There is also disagreement about the value of the constant $A$
\cite{Son:1999cm,Beane:2000ms,Manuel:2000wm,Rho:2000xf,Hong:2000ei}.
The results given in \cite{Beane:2000ms} and \cite{Son:1999cm} 
are, respectively
\be
\label{a_res}
A=\frac{\Delta \bar\Delta}{4 \pi^2}\log(\mu/\Delta), \hspace{1cm}
A=\frac{3 \Delta^2}{4 \pi^2}.
\ee
Using the first of these two results a $K^0$ condensate forms if
\be
\label{mscrit}
m_s^3 > \left(\frac{144}{21-8\log 2}\right)
        m_u\Delta\bar\Delta \log(\mu/\Delta) .
\ee
In Fig. \ref{fig_kaon_ms} we show the dependence of the kaon 
mass on $m_s$ for $p_F=500$ MeV and with $\Delta$ and $f_\pi$
calculated to leading order in perturbation theory. We observe
that the $K^0$ becomes massless for $m_s\simeq 60$ MeV. There
is obviously some uncertainty associated with the use of first 
order perturbation theory. An estimate of this uncertainty is 
provided by the scale dependence of the result. We have 
calculated $m_K$ with $g$ evaluated at the scale $\Lambda
=p_F$. Varying $\Lambda$ between $p_F/2$ and $2p_F$ gives
critical strange quark masses between 39 MeV and 67 MeV.

 If charge neutrality is enforced we have to add the contribution 
of electrons to the thermodynamic potential, $\Omega(\Sigma,\mu_e)= 
\Omega_{GB}(\Sigma,\mu_e)-\mu_e^4/(12\pi^2)$. The ground state is 
determined by minimizing $\Omega$ with respect to $\Sigma$ subject 
to the condition that $\partial\Omega/(\partial\mu_e)=0$. In the 
isospin symmetric limit these conditions are satisfied by pure 
$K^0$ condensation with $\alpha$ as determined above and $\sin(
\theta_1)=\sin(\theta_2)=1$. This conclusion remains valid
in the case $m_d>m_u$ because the light quark mass difference 
also disfavors $K^+$ condensation compared to $K^0$ condensation.  

 The effect of a small electron chemical potential can also be read 
off from Eq.~(\ref{leff2}). A positive electron chemical potential lowers 
the energy of negatively charged Goldstone modes and increases the energy 
of positively charged modes,
\be 
\label{mgb}
 E_{\pi^\pm} = \pm \mu_e + m_{\pi^\pm}, \hspace{1cm}
 E_{K_\pm}   = \pm \mu_e + m_{K_\pm}.
\ee
A meson condensate will form when $\mu_e$ equals the mass of the 
lightest negatively charged state. Let us again consider the limit
of exact isospin symmetry, $m_u=m_d=m$. The mass of the $K^-$
is $m_{K^-}=(2\sqrt{A}/f_\pi)\sqrt{mm_s}+m_s^2/(2p_F)$ and the 
mass of the $\pi^-$ is $m_{\pi^-}=\sqrt{2}(2\sqrt{A}/f_\pi)\sqrt{mm_s}$.
For very small $m_s$ the lightest negatively charged particle is the 
$K^-$, but for $m_s^2/(2p_F) >(\sqrt{2}-1)(2\sqrt{A}/f_\pi)\sqrt{mm_s}$ 
the lightest negative state is the $\pi^-$. For 
negative electron chemical potentials a $K^+$ condensate is always 
favored. We should note that the masses of charged Goldstone bosons
are modified by electromagnetic effects. The electromagnetic self
energy in the CFL phase was estimated to be $m^2_{em}\sim \alpha_{em}
\Delta^2$ \cite{Hong:2000ng,Manuel:2001xt}. At sufficiently large 
baryon density this effect will dominate over the $O(M^2)$ 
contribution to the Goldstone boson masses.

%%%%%%%%%%%%%%%%%%%%%%%%%%%%%%%%%%%%%%%%%%%%%%%%%%%%%%%%%%%%%%%%%%%%%%%%%
\section{Matching calculation for the $O(M^4)$ terms} 
\label{sec_match}
%%%%%%%%%%%%%%%%%%%%%%%%%%%%%%%%%%%%%%%%%%%%%%%%%%%%%%%%%%%%%%%%%%%%%%%%%

 In the weak coupling regime the coefficients appearing in the 
Lagrangian Eq.~(\ref{leff2}) can be computed by matching to
perturbative QCD. In this section we will perform the matching 
calculation for the $M^4$ terms in Eq.~(\ref{leff2}). Our goal 
is twofold: to strengthen and illustrate the symmetry arguments 
presented in the previous section and to clarify the calculations
of $f_\pi$ in the literature\footnote{We thank D.~Kaplan for 
suggesting this calculation to us.}.

\begin{figure}
\begin{center}
\leavevmode
\vspace{1cm}
\epsfxsize=10cm
\epsffile{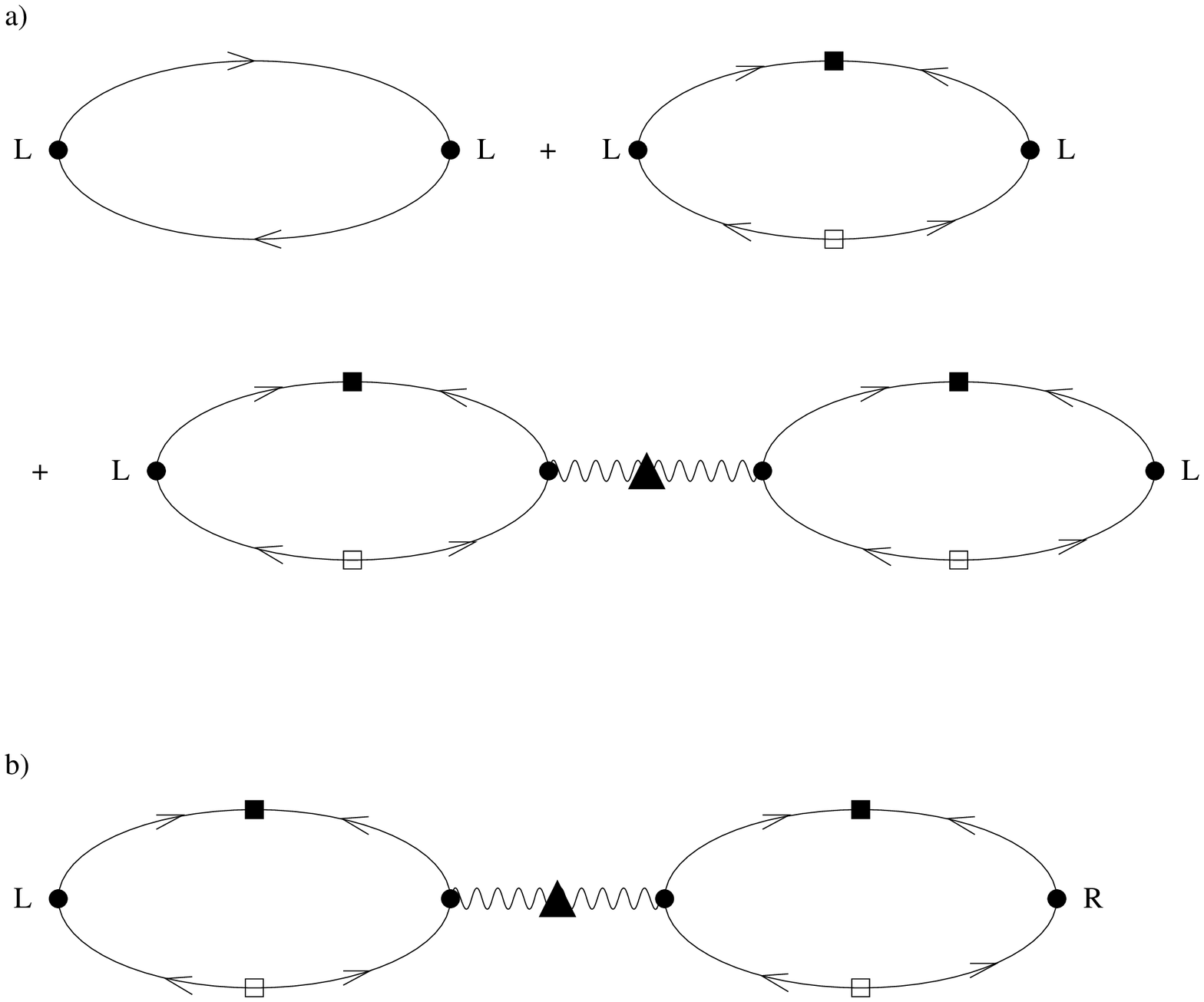}
\end{center}
\begin{center}
\leavevmode
\vspace{1cm}
\epsfxsize=10cm
\epsffile{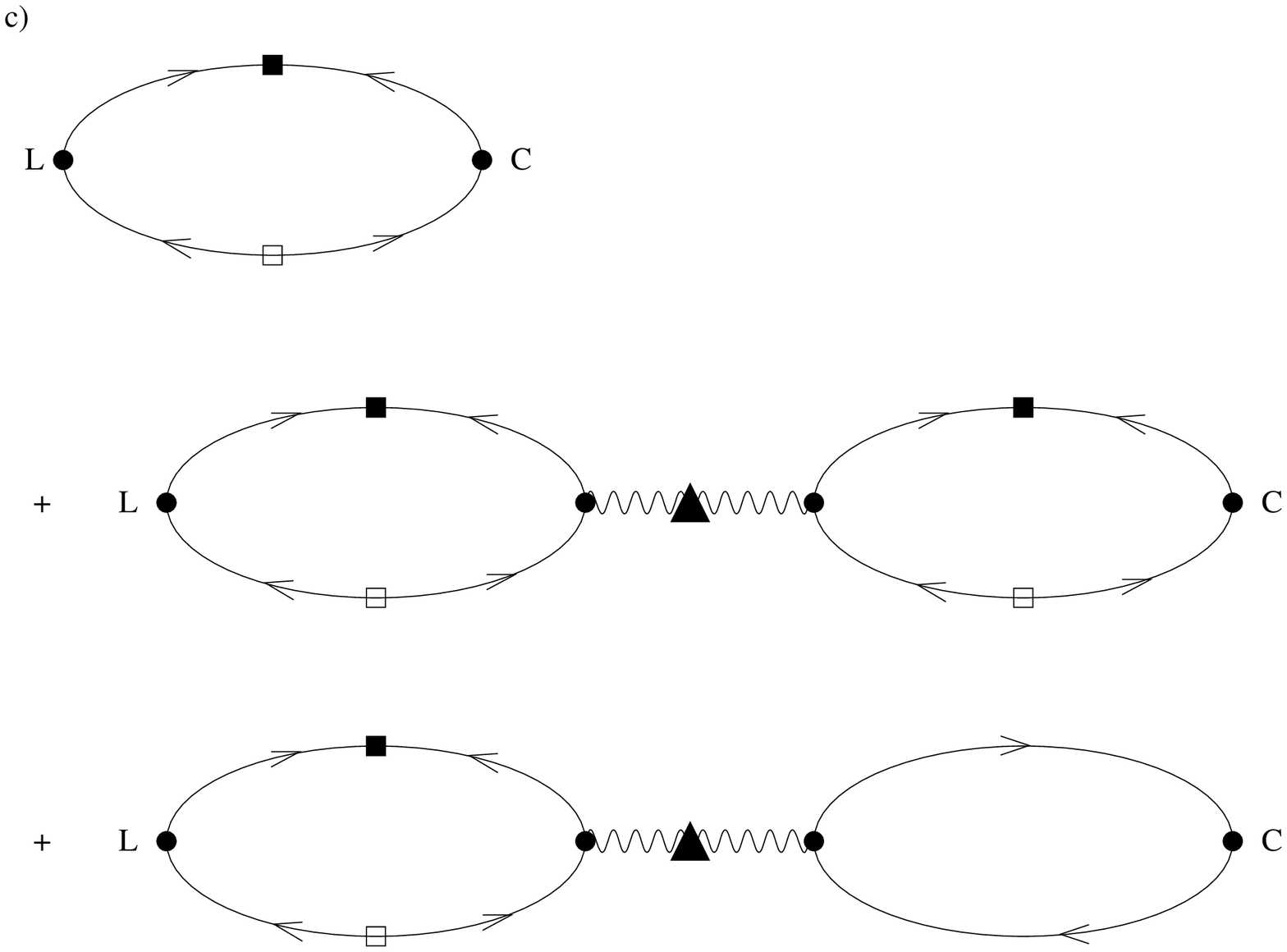}
\end{center}
\caption{\label{fig_fpi}
Diagrams contributing to the two-point functions of two $L$ 
currents (Fig. a), one $L$ and one $R$ current (Fig. b), and
one $L$ and one color current (Fig. c). The squares denote the 
anomalous fermion self energy while the triangle denotes a 
resummed gluon propagator.}
\end{figure}

 We begin by calculating the one-loop polarization functions
for the zeroth component of left-handed flavor currents 
$j_L$, right-handed flavor currents $j_R$ and (transposed) color 
currents $j_c^T$. In the limit $\omega=0, k\to 0$ we find 
\be
\label{pol_1l}
\Pi_{00}^{AB}(0) = -\left( \begin{array}{rrr}
  \frac{1}{2} &      0      & -\frac{1}{2} \\
         0    & \frac{1}{2} & -\frac{1}{2} \\
 -\frac{1}{2} &-\frac{1}{2} &    1 
\end{array}\right) m_D^2 ,
\ee
where the indices $A,B$ correspond to $(j_L,j_R,j_c^T)$ and
we have introduced the quantity
\be
 m_D^2 = \frac{21-8\log(2)}{18}
  \left( \frac{\mu^2}{2\pi^2}\right),
\ee 
which is, up to a factor $g^2$, the Debye mass 
\cite{Son:1999cm,Rischke:2000ra}. The $LL$ and $RR$ components 
of (\ref{pol_1l}) receive contributions both from diagrams with 
normal propagators and from diagrams with anomalous propagators, 
see Fig.~\ref{fig_fpi}. The $LC$ and $RC$ components only receive 
contributions from diagrams with anomalous propagators
\cite{Manuel:2001xt}. The overall coefficient is nevertheless
exactly the same. The $CC$ entry is twice bigger than the
$LL$ and $RR$ entries because it receives contributions
from both left and right handed fermions. 

 The matrix (\ref{pol_1l}) is not diagonal, so there is mixing
between gluons and left or right handed flavor currents. Also,
there is no mixing between left and right handed flavor currents,
contrary to what we would expect for a system with broken chiral
symmetry. These defects can be cured by re-summing bubble chains 
with intermediate gluons. In practice we only have to compute 
the two-loop contribution because higher order diagrams 
simply correspond to replacing the free gluon propagator 
$1/(\omega^2-k^2)$ with the dressed propagator $1/(\omega^2
-k^2-g^2m_D^2)$. The two-loop contributions to the polarization
function are superficially suppressed by a factor $g^2$ but
in the limit $\omega,k\to 0$ the factor $g^2$ in the numerator
is canceled by the screening mass $g^2m_D^2$ in the denominator. 

 Summing all bubble chains we get
\be
\label{pol_2l}
\Pi^{AB}_{00}(0) = -\left( \begin{array}{rrr}
  \frac{1}{4} &-\frac{1}{4} &    0 \\
 -\frac{1}{4} & \frac{1}{4} &    0 \\
        0     &     0       &    1 
\end{array}\right) m_D^2.
\ee
We observe that flavor and color currents are decoupled and 
that the mixing matrix between left and right handed current 
has the form expected for a system with broken chiral symmetry. 
To leading order in $g^2$ there are no additional contributions
to the polarization function in the soft limit. We can now 
match the result (\ref{pol_2l}) against the low energy theory
\be
\label{nls_g}
{\cal L} = \frac{f_\pi^2}{4}
  {\rm Tr}(\nabla_0\Sigma\nabla_0\Sigma^\dagger),
\ee
where the covariant derivative $\nabla_0\Sigma=\partial_0\Sigma+i
W_L\Sigma-i\Sigma W_R$ determines the coupling to left and right
handed gauge fields $W_{L,R}$. Matching the gauge field mass terms
against (\ref{pol_2l}) gives $f_\pi^2=m_D^2$, which is the result
of Son and Stephanov \cite{Son:1999cm,Zarembo:2000pj,Miransky:2001bd}. 

 This  result can also be obtained in a different way. Since
the gluon field acquires a large mass of order $g\mu\gg\Delta$
it does not appear in the low energy effective theory and we
should be able to integrate it out \cite{Casalbuoni:1999wu}. 
The matrix in (\ref{pol_1l}) has eigenvalues $\lambda=-1/2,-3/2,0$ 
and eigenvectors $(1,-1)/\sqrt{2}$, $(1,1,-2)/\sqrt{6}$ and 
$(1,1,1)/\sqrt{3}$. The vanishing eigenvalue corresponds to 
the generators of the unbroken $SU(3)_{L+R+C}$. The one-loop 
polarization function can be matched against the following 
mass term for the gauge fields
\be
\label{wm1}
{\cal L} = \frac{m_D^2}{4}\left[ 
  \frac{1}{2} (W_L-W_R)^2 + 
  \frac{1}{2} (W_L+W_R-2A_0^T)^2) \right] .
\ee
The gauge field mass term still has the structure $\frac{1}{2}
(\frac{m_D^2}{2})(W_L^2+W_R^2+{\rm mixing})$ apparent in (\ref{pol_1l}). 
Integrating out the gluon field $A_0$ eliminates the second term in 
(\ref{wm1}) and we are left with 
\be
\label{wm2}
{\cal L} = \frac{m_D^2}{4} \, \frac{1}{2} (W_L-W_R)^2,
\ee
which has the structure expected from the low energy 
effective theory (\ref{nls_g}). Matching (\ref{wm2})
against (\ref{nls_g}) gives $f_\pi^2=m_D^2$ as before.
The important point is that in both approaches, summing
bubble chains or integrating out the gluon field at tree
level, the mixing between flavor and color currents
cuts down the coefficient of the quadratic terms $W_L^2$
and $W_R^2$ by a factor of 2 and introduces mixing between 
left and right handed currents.
  
  We are now in a position to perform the matching calculation
for the $M^4$ term in Eq.~(\ref{leff2}). In App.~\ref{app_int} 
we present an alternative argument based on integrating out
the gauge field. We are concerned with a possible mass term 
of the form 
\bea
{\cal L} &=& -\frac{\bar{f}^2}{4}
  {\rm Tr}\left[ (MM^\dagger\Sigma-\Sigma M^\dagger M)
    (M^\dagger M\Sigma^\dagger-\Sigma^\dagger MM^\dagger)\right]
   \\
\label{lm4}
 &=& -\frac{\bar{f}^2}{2}
  {\rm Tr}\left[ (MM^\dagger\Sigma M^\dagger M\Sigma^\dagger
  - MM^\dagger MM^\dagger) \right] .
\eea 
We will determine $\bar f$ by computing the shift in the ground 
state energy proportional to ${\rm Tr}[MM^\dagger M^\dagger M]$ 
and ${\rm Tr}[(MM^\dagger)^2]$ in both QCD and in the effective 
theory. In the effective theory the shift is given by
\be
\label{matchqcd}
\Delta {\cal E} = \frac{\bar f^2}{2} 
 {\rm Tr}\left[(MM^\dagger)(M^\dagger M)-(MM^\dagger)^2\right].
\ee
We note that the two terms in Eq.~(\ref{lm4}) can be distinguished
even in the phase $\Sigma=1$ by the the relative position of $M$  
and $M^\dagger$. We also note that other $O(M^4)$ terms allowed 
by the symmetries of QCD give structures that are different from 
the ones that appear in Eq.~(\ref{matchqcd}). 

\begin{figure}
\begin{center}
\leavevmode
\vspace{1cm}
\epsfxsize=10cm
\epsffile{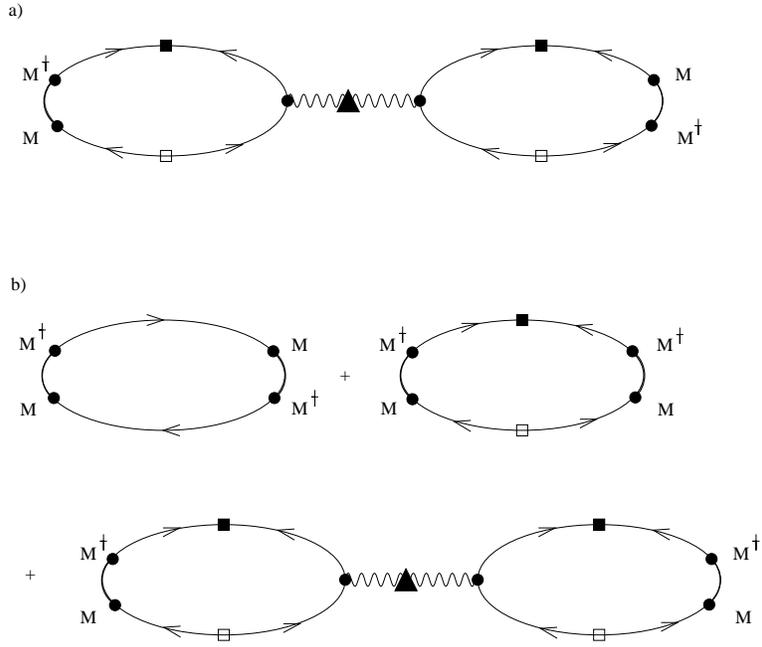}
\end{center}  
\caption{\label{fig_m4}
Fig.~a) shows the diagram in the microscopic theory which is matched 
against the $MM^\dagger \Sigma M^\dagger M\Sigma$ term in the chiral 
theory. Fig.~b) shows the diagrams which are matched against the 
$(MM^\dagger)^2$ term.}
\end{figure}

 In the microscopic theory the shift in the vacuum energy 
proportional to  ${\rm Tr}[MM^\dagger M^\dagger M]$ 
and ${\rm Tr}[(MM^\dagger)^2]$ comes from the graphs in 
Figs.~\ref{fig_m4}a) and b). The ${\rm Tr}[MM^\dagger M^\dagger M]$ 
term is given by
\bea 
\Delta{\cal E} &=&  \frac{1}{(2p_F)^2}\left(\frac{m_D^2}{2}\right)
   {\rm Tr}\left[MM^\dagger\lambda^a\right]\cdot
    \frac{\delta^{ab}}{m_D^2} \cdot
  \left( \frac{m_D^2}{2}\right) {\rm Tr}\left[M^\dagger M\lambda^b\right]
  \nonumber \\
 &=& \frac{m_D^2}{2}\frac{1}{(2p_F)^2}
 {\rm Tr}\left[(MM^\dagger)(M^\dagger M)\right],
\eea
and the ${\rm Tr}[(MM^\dagger)^2]$ term is 
\be
\Delta{\cal E} = -\frac{m_D^2}{2}\frac{1}{(2p_F)^2}
 {\rm Tr}\left[(MM^\dagger)^2\right].
\ee
Matching these results against Eq.~(\ref{matchqcd}) we conclude that 
\be
 \bar{f}^2 = \frac{f_\pi^2}{(2p_F)^2} ,
\ee
which is the result we derived in section \ref{sec_3flavor} from making 
the time derivate covariant with respect to time dependent flavor 
transformations.

%%%%%%%%%%%%%%%%%%%%%%%%%%%%%%%%%%%%%%%%%%%%%%%%%%%%%%%%%%%%%%%%%%%%%%%%%
\section{Linear Response}
\label{sec_resp}
%%%%%%%%%%%%%%%%%%%%%%%%%%%%%%%%%%%%%%%%%%%%%%%%%%%%%%%%%%%%%%%%%%%%%%%%%

 In this section we offer a different perspective on the results 
discussed in the previous sections by using linear response theory. 
We shall also provide a more microscopic explanation of why the two 
and three flavor cases behave so differently. In the three flavor
case the system responds to a non-zero electron chemical potential
by forming a condensate of collective excitations. In the two flavor
case, on the other hand, the response is carried only by the 
ungapped fermions. From an effective field theory point of view 
this is simply due to the fact that three flavor CFL quark matter 
has broken chiral symmetry and the low energy effective description 
contains charged collective modes whereas the two flavor theory 
has unbroken chiral symmetry and the low energy theory contains 
ungapped fermions and neutral modes. 

   In order to set the stage for the discussion of superfluid quark 
matter we briefly review the response of ordinary quark matter. The 
grand canonical potential of non-interacting quarks at zero temperature
is given by
\be
\label{Omega}
 \Omega \,=\,-p \,=\, -\frac{N_c}{12\pi^2}\sum_f
   \left[ \mu_fk_f\left(\mu_f^2-\frac{5}{2}m_f^2\right)
   + \frac{3}{2}m_f^4 \log\left(\frac{\mu_f+k_f}{m_f}\right)\right],
\ee
with $k_f=\sqrt{\mu^2_f-m_f^2}$ is the Fermi momentum and and $\mu_f$ 
the chemical potential for the quark flavor $f=u,d,s$. The quark density 
is given by
\be
 n_f\,=\, -\frac{\partial\Omega}{\partial\mu_f}
 \, =\, \frac{N_c k_f^3}{3\pi^2}
\ee
It is convenient to decompose the chemical potential into
baryon charge, isospin, and hypercharge components
\bea
 \mu_{u} &=& \mu + \frac{1}{2}\mu_I +\frac{1}{2\sqrt{3}}\mu_Y, \\
 \mu_{d} &=& \mu - \frac{1}{2}\mu_I +\frac{1}{2\sqrt{3}}\mu_Y, \\
 \mu_{s} &=& \mu -\frac{1}{\sqrt{3}}\mu_Y. 
\eea
We also note that $\mu_I=\sqrt{3}\mu_Y=-\mu_e$ acts like a chemical 
potential for electric charge. We can now study the response of the 
system to an external chemical potential or a change in the quark 
masses. We begin with the flavor symmetric case $m_u=m_d=m_s=0$. The 
isospin and hypercharge susceptibilities are
\be 
\label{suscep}
\chi_I = \frac{\partial n_I}{\partial\mu_I} =
-\frac{\partial^2\Omega}{\partial\mu_I^2} =
\chi_Y = \frac{\partial n_Y}{\partial\mu_Y} = 
-\frac{\partial^2\Omega}{\partial\mu_Y^2} =
 N_c \left(\frac{\mu^2}{2\pi^2}\right).
\ee
This result has a very simple interpretation. The change
in the isospin or hypercharge density as a function of the 
corresponding chemical potential is simply given by the 
density of states on the Fermi surface. The susceptibilty
(\ref{suscep}) can also be calculated in a different 
way, using the fact that $\chi$ is the flavored vector
current correlation function at zero momentum. We have
\be
\label{pi00}
\chi_I = -\Pi_{I}(\omega\!=\!0,\vec{k}\to 0) =
-\int d^4x\,\langle j_0^3(x)j_0^3(0)\rangle
\ee
with $j_\mu^a(x)=\bar\psi(x)\gamma_\mu\frac{\tau^a}{2}\psi$.
The correlation function (\ref{pi00}) has a vacuum piece
and a density dependent piece. The density dependent piece
is dominated by the contribution of particles and holes
in the vicinity of the Fermi surface. We can calculate this
contribution using the effective theory proposed in 
\cite{Hong:2000tn,Hong:2000ru}. We get
\be
\chi_I = \lim_{\omega,k\to 0}
 N_c \int\frac{d^4p}{(2\pi)^4}
 \frac{1}{(p_0-\epsilon_p)(p_0+\omega-\epsilon_{p+k})}
 = N_c \int \frac{d^3p}{(2\pi)^3} 
        \frac{\partial n}{\partial \epsilon}
 =  N_c\left(\frac{\mu^2}{2\pi^2}\right),
\ee
where $\epsilon_p= E_p-\mu$, $E_p=\sqrt{p^2+m^2}$, and
$n(\epsilon)$ is the density of states. This result 
obviously agrees with Eq.~(\ref{suscep}). 

 From the grand canonical potential (\ref{Omega}) we can
also determine the response of the system to non-zero
quark masses. The derivative of the hypercharge density
with respect to the strange quark mass is given by
\be
\label{suscep_ms}
 \left. \mu\frac{\partial n_Y}{\partial m_s^2} \right|_{m_s^2=0}
 =-\left. \mu\frac{\partial^2 \Omega}{\partial m_s^2\partial \mu_Y}
 \right|_{m_s^2=0}= \frac{N_c}{\sqrt{3}} 
\left(\frac{\mu^2}{2\pi^2}\right).
\ee
This result expresses the simple fact that the number of
strange quarks is depleted compared to the number of 
non-strange quarks as the mass of the strange quark is 
increased. Again, we can compute this susceptibility 
using diagrammatic techniques. Computing a one-loop
graph with one insertion of $\mu_Y$ and one insertion
of $m_s^2/(2\mu)$ we reproduce (\ref{suscep_ms}). 

 When we study real physical systems we are interested 
in the response of the system subject to the constraint
that certain quantities are exactly conserved. In the case 
of neutron stars, for example, we are interested in the 
composition of quark matter subject to the condition that 
the baryon density is fixed and the net density of electric 
charge is zero. For this purpose we consider the thermodynamic 
potential as a function of the quark density  $\rho_q=3\rho_B$,
the up and down quark fractions $x=\rho_u/\rho_q$ and $y=\rho_d/
\rho_q$, and the electron chemical potential $\mu_e$
\bea
\omega(\rho_q,x,y,\mu_e) &=& F(\rho_q,x,y)-\mu_e Q
 =  \frac{3\pi^{2/3}}{4}
   \rho_q^{4/3} \left\{ x^{4/3} + y^{4/3}
  + (1-x-y)^{4/3} \right. \nonumber \\
& & \mbox{}\left. 
 + \pi^{-4/3}\rho_q^{-2/3}m_s^2
  (1-x-y)^{2/3} \right\} 
  + \mu_e \rho_q \left(x-\frac{1}{3}\right)
     - \frac{1}{12\pi^2}\mu_e^4 .
\label{eps}
\eea
We have neglected higher order terms in the strange quark 
mass as well as the mass of the electron. In order to determine 
the ground state we have to make (\ref{eps}) stationary with 
respect to $x,y,\mu_e$. Minimization with respect to $x$ and $y$ 
enforces $\beta$ equilibrium, while minimization with respect 
to $\mu_e$ ensures charge neutrality. We find 
\be
 \mu_e\simeq \frac{m_s^2}{4p_F},
\ee
which shows that there is a small non-zero $\mu_e$ and
a corresponding suppression of strange quarks with respect 
to light quarks even at high density. 

 We would now like to study how these results are modified
in superfluid phases of QCD. We begin with a simple toy model 
introduced by Rajagopal and Wilczeck \cite{Rajagopal:2000ff}.
The model contains two quark flavors, up and down, that 
pair in a spin singlet state which is anti-symmetric in 
both color and flavor. The pair condensate is described
by the order parameter $\langle \epsilon^{ab}u^a C\gamma_5
d^b\rangle$. Here, $a,b$ are color indices that only take
on the values 1 and 2. One may think of this toy model as
$N_f=2$ QCD where the contribution of the third, unpaired, 
quark color is ignored. Alternatively, we may think of this
theory as $N_c=2$ QCD.

 We can calculate the response in the superfluid in the same 
way we did in the normal phase, using the relation between the 
quark number susceptibilities and the $00$-component of the 
polarization function. In the superfluid phase there are two 
contributions, coming from the normal and anomalous components 
of the quark propagator. For the quark number susceptibility 
we get
\be
 \chi_B = -\Pi_{00}(\omega=0,\vec{k}\to 0) = 
  4N_c \int\frac{d^4p}{(2\pi)^4} \left\{ 
 \frac{p_0^2+\epsilon_p^2}{(p_0^2-\epsilon_p^2-\Delta^2)^2}
 - \frac{\Delta^2}{(p_0^2-\epsilon_p^2-\Delta^2)^2}
  \right\} ,
\ee
where the first term is the contribution from the 
normal quark propagator and the second term is the 
anomalous contribution. The two contributions are 
exactly equal and sum up to 
\be
 \chi_B = 4N_c \left\{ \left(\frac{\mu^2}{4\pi^2}\right)
  + \left(\frac{\mu^2}{4\pi^2}\right) \right\} 
 = 4N_c \left(\frac{\mu^2}{2\pi^2}\right),
\ee
which is equal to the result in the normal phase. We should note that 
the first term alone only contributes half the susceptibility 
in the normal phase, even though the susceptibility is independent 
of the gap and the naive $\Delta\to 0$ limit of the first graph 
would seem to correspond to the susceptibility in the normal phase. 
This is due to the fact that the $\omega\to 0$ and $\Delta\to 0$ 
limits do not commute. This phenomenon is well known from calculations 
of the screening mass in other many body systems \cite{Abrikosov:1963}.

 The calculation of the isospin susceptibility proceeds
along exactly the same lines, only the isospin factors 
of the two diagrams are different. The isospin factor
of the normal contribution is ${\rm tr}[\tau_3\tau_3]=2$,
while the isospin factor of the second term is 
${\rm tr}[\tau_3\tau_2\tau_3\tau_2]=-2$. The two 
contributions cancel exactly and the isospin susceptibility
is zero. This results has a simple physical interpretation. 
The superfluid order parameter in $N_f=N_c=2$ QCD is a flavor 
singlet and the only broken symmetry is the $U(1)$ of baryon 
number. As a result there is only one massless state, the 
$U(1)$ Goldstone boson. This state couples to the baryon 
density and leads to a non-zero baryon number susceptibility
but it does not couple to isospin. All states that carry 
isospin have energies of the order of the gap, so $\chi_I$
remains zero as long as $\mu_I<\Delta$. 

 We can also see how the calculation of the isospin susceptibility 
differs in the case of CFL quark matter. Because of the symmetries 
of the CFL phase there are two types of quasi-particles, an $SU(3)$
octet with gap $\Delta_8=\Delta$ and an $SU(3)$ singlet with gap
$\Delta_1=2\Delta$. Up to degeneracy factors the two types of
quasi-particles contribute equally to the quark number susceptibility.
We find $\chi_B= 18\mu^2/(2\pi^2)$ which is equal to the result in 
the normal phase. The calculation of the isospin susceptibility
is more complicated. We get
\bea
 \chi_I &=& 
 2 \int\frac{d^4p}{(2\pi)^4} \left\{ 
 \frac{7}{6}\frac{p_0^2+\epsilon_p^2}
        {(p_0^2-\epsilon_p^2-\Delta_8^2)(p_0^2-\epsilon_p^2-\Delta_8^2)}
+\frac{1}{3}\frac{p_0^2+\epsilon_p^2}
        {(p_0^2-\epsilon_p^2-\Delta_8^2)(p_0^2-\epsilon_p^2-\Delta_1^2)}
\right.\nonumber\\
\label{nf3_chiv}
 & & \left.\hspace{1cm}\mbox{}
 - \frac{1}{3}\frac{\Delta_8^2}
        {(p_0^2-\epsilon_p^2-\Delta_8^2)(p_0^2-\epsilon_p^2-\Delta_8^2)}
 - \frac{1}{3}\frac{\Delta_8\Delta_1}
        {(p_0^2-\epsilon_p^2-\Delta_8^2)(p_0^2-\epsilon_p^2-4\Delta_1^2)}
  \right\}.
\eea
The first term comes from particle-hole diagrams with two 
octet quasi-particles while the second term comes from diagrams
with one octet and one singlet quasi-particle. There is no
coupling of an octet field to two singlet particles. The third
and fourth term are the corresponding contributions from 
particle-particle and hole-hole pairs. The four integrals in 
(\ref{nf3_chiv}) give
\be
 \chi_I = 2\left\{ \frac{7}{6} +\frac{1}{3}
  - \frac{1}{3} - \frac{4\log(2)}{9}\right\}
   \left(\frac{\mu^2}{4\pi^2}\right)
 = \frac{21-8\log(2)}{18} \left( \frac{\mu^2}{2\pi^2} \right)
 \simeq 0.86 \left( \frac{\mu^2}{2\pi^2} \right),
\ee
which should be compared to $\chi_I=3\mu^2/(2\pi^2)$ in the 
normal phase. We observe that there is a partial cancellation
between the normal and anomalous contributions. However, 
because of the more complicated flavor structure this 
cancellation is not exact. The isospin density induced by 
an isospin chemical potential is reduced by a factor 
$\sim 3.5$ compared to the normal phase, but it does not 
vanish. In linear response theory we expand around the 
point $\mu_I=m_u=m_d=m_s=0$. In the real world the quark 
masses are non-zero and there is a critical isospin chemical 
potential $\mu_I^{crit}\neq 0$ below which the isospin 
susceptibility vanishes. In order to see a threshold 
behavior in $\mu_I$ we have to resum mass corrections. 
This is most efficiently accomplished using the effective 
chiral description developed in section \ref{sec_3flavor} , 
see equation (\ref{k0+_V}).

%%%%%%%%%%%%%%%%%%%%%%%%%%%%%%%%%%%%%%%%%%%%%%%%%%%%%%%%%%%%%%%%%
\section{Summary}
\label{sec_sum}
%%%%%%%%%%%%%%%%%%%%%%%%%%%%%%%%%%%%%%%%%%%%%%%%%%%%%%%%%%%%%%%%%

 We have studied the response of three flavor quark matter to
a non-zero electron chemical potential and a non-zero strange
quark mass. We have focussed on the regime $\mu_e,m_s^2/(2p_F)
<\Delta$ in which the perturbation does not destroy color-flavor
locking. We have identified a new scale $\mu_e,m_s^2/(2p_F)\sim
\sqrt{m_{u,d}m_s}(\Delta/p_F)$ which corresponds to the onset 
of pion or kaon condensation \cite{Migdal:1973,Sawyer:1972,Scalapino:1972,Kaplan:1986,Brown:1987,Politzer:1991}.
This scale is parametrically much smaller than the gap. If 
CFL quark matter exists in the core of a neutron star it is 
likely to be $K^0$ condensed. Both with or without a kaon 
condensate there are no electrons present \cite{Rajagopal:2000ff}.
If CFL quark matter is in contact with a hadronic phase that supports 
a large electron chemical potential the surface layer is likely to be 
$K^-$ or $\pi^-$ condensed \cite{Alford:2001zr}. 
 
 These results are based on an analysis of how to incorporate 
$\mu_e$ and $m_s^2/(2p_F)$ in the chiral effective theory. Both terms 
enter as constant flavor gauge fields, with coefficients completely 
determined by $f_\pi$. The contribution of the $m_s^2/(2p_F)$
term to the Goldstone boson masses is of higher order in the 
quark masses as compared to the leading order $\sqrt{mm_s}
(\Delta/p_F)$ term. It can nevertheless become dominant 
because the $O(m)$ term is suppressed by powers of $\sqrt{m/m_s}$
and $(\Delta/p_F)$. As a consequence the $O(m^2)$ term can cancel
the $O(m)$ term without leading to a breakdown of the low
energy expansion.

Acknowledgements: We would like to thank S.~Beane, D.~Kaplan, 
C.~Manuel, K.~Rajagopal, S.~Reddy, M.~Savage, D.~Son, M.~Stephanov 
and D.~Toublan for useful discussions. The work of T.~S.~was 
supported in part by US DOE grant DE-FG-88ER40388. The work 
of P.~B.~was supported in part by the Director, Office
of Energy Research, Office of High Energy and Nuclear Physics, 
Division of Nuclear Physics, and by the Office of Basic Energy
Science, Division of Nuclear Science, of the U.S.  Department 
of Energy under Contract No. DE-AC03-76SF00098.

\newpage      
\appendix

%%%%%%%%%%%%%%%%%%%%%%%%%%%%%%%%%%%%%%%%%%%%%%%%%%%%%%%%%%%%%%%%%%%%%%%%%
\section{Mass terms induced by the color symmetric diquark condensate}
\label{app_sym}
%%%%%%%%%%%%%%%%%%%%%%%%%%%%%%%%%%%%%%%%%%%%%%%%%%%%%%%%%%%%%%%%%%%%%%%%%

  The $O(M^2)$ mass term in (\ref{leff2}) gives anomalously small
Goldstone boson masses of the order $m_{GB}\sim \sqrt{m m_s}
(\Delta/p_F)$. We already noted that mass terms not suppressed 
by $(\Delta/p_F)$ cannot appear at $O(M^2)$. For strange mesons
the $O(M^2)$ mass term also contains an additional suppression
factor $\sqrt{m/m_s}$. Here, $m$ is the mass of the light quarks and 
$m_s$ is the strange quark mass. The fact that all Goldstone
boson masses are proportional to the light quark mass is related
to the fact that the CFL order parameter is totally anti-symmetric
in flavor. This flavor structure also leads to an accidental 
symmetry of the effective theory at $O(M^2)$. If $m_s=0$ but 
$m\neq 0$ we find an octet of exact Goldstone bosons, even though 
the unbroken flavor symmetry is only $SU(2)$. 

 There are mass terms at $O(M^2)$ that are consistent with the
symmetries of the CFL phase that will remove the accidental 
symmetry and give contributions to the kaon mass that are
proportional to $m_s(\Delta/p_F)$. These terms are induced
by the color-flavor symmetric gap parameter
\be
\label{del_sym}
 \Delta^{ab}_{ij} = \Delta_S \left(\delta^a_i\delta^b_j
 + \delta^a_j\delta^b_i \right).
\ee
The symmetric gap is consistent with the symmetries of the 
CFL phase but disfavored by the interaction. In particular,
one-gluon exchange is repulsive in the color-symmetric 
quark-quark channel. In perturbative QCD, a small symmetric
gap is generated by mixing with the primary gap parameter. 
We find \cite{Schafer:1999fe}
\be 
\label{s-gap}
 \Delta_S = \frac{g}{\pi}\frac{\sqrt{2}\log(2)}{36}\Delta_A,
\ee
where $\Delta_A$ is the color-flavor anti-symmetric gap
parameter. 

 We can calculate the contribution of $\Delta_S$ to the Goldstone 
masses using the methods of Beane et al.~\cite{Beane:2000ms}. 
Including the effects of both $\Delta_A$ and $\Delta_S$ 
we find
\bea
{\cal L} &=& -\frac{\Delta_A\bar\Delta_A}{4\pi^2}
 \log\left(\frac{\Delta_A}{p_F}\right)
 \left( {\rm Tr}(M\Sigma){\rm Tr}(M\Sigma)-
        {\rm Tr}(M\Sigma M\Sigma) + {\rm h.c.} \right) 
  \nonumber \\
 & & \mbox{}
-\frac{\Delta_S\bar\Delta_S}{4\pi^2}
 \log\left(\frac{\Delta_S}{p_F}\right)
 \left( {\rm Tr}(M\Sigma){\rm Tr}(M\Sigma)+
        {\rm Tr}(M\Sigma M\Sigma) + {\rm h.c.} \right). 
\eea
Here, $\bar\Delta_{A,S}$ are the flavor anti-symmetric and
symmetric ``anti-gaps''. For the purpose of estimating the 
relative size of the two mass terms we shall assume that
$\bar{\Delta}_{A,S}\simeq \Delta_{A,S}$. We can now 
calculate the correction to the charged kaon mass, 
\bea
 m_{K^\pm} &=& \left[ \frac{4A_A}{f_\pi^2} m_d(m_u+m_s)
     + \frac{4A_S}{f_\pi^2} (m_u+m_s)(2m_s+2m_u+m_d)\right]^{1/2} 
        \nonumber \\
 &\simeq & \frac{2\sqrt{A_A}}{f_\pi}\sqrt{m m_s} \left(
  1 + \left(\frac{\Delta_S}{\Delta_A}\right)^2 \frac{m_s}{m_u} +\ldots
 \right) , 
\eea
with $A_A=\Delta_A^2/(4\pi^2)\log(p_F/\Delta_A)$. Using
(\ref{s-gap}) we observe that even for $m_s/m_u\simeq 20$
the correction to the kaon mass due to the color-symmetric gap 
is still small. 

%%%%%%%%%%%%%%%%%%%%%%%%%%%%%%%%%%%%%%%%%%%%%%%%%%%%%%%%%%%%%%%%%%%%%%%%%
\section{$\mu Q +MM^\dagger/(2p_F)$ terms from integrating out the 
gauge field}
\label{app_int}
%%%%%%%%%%%%%%%%%%%%%%%%%%%%%%%%%%%%%%%%%%%%%%%%%%%%%%%%%%%%%%%%%%%%%%%%%

 Following the discussion in section \ref{sec_match}
we can also derive the $O(M^4)$ terms by integrating
out the gauge field. This discussion will also make
it clear that the $M^\dagger M$ and $MM^\dagger$ terms
enter in the effective lagrangian like gauge fields,
together with flavor non-singlet chemical potentials.

 In this section we would also like to show how, by 
explicitly keeping track of the orientation of the 
CFL order parameter, we can determine how the chiral
field $\Sigma$ enters into the mass terms. This is 
useful because at higher order the number of independent
terms in the chiral lagrangian quickly proliferates
and it becomes more difficult to identify the diagrams
in the microscopic theory that correspond to a given term
in the effective lagrangian. 

 In order to match the microscopic theory against
the effective theory in the vacuum ($\Sigma=1$) phase 
we calculate diagrams in the microscopic theory using
the Nambu-Gorkov propagators in the normal CFL phase. The 
inverse Nambu-Gorkov propagator for the $\psi_+$ field
is given by
\be 
\label{sinv}
S^{-1} = \left(\begin{array}{cc}
   p_0-\epsilon_p  & \Delta \\
   \Delta  & p_0+\epsilon \end{array}\right),
\ee
with the anomalous self energy 
\be
\label{del_cfl}
 (\Delta_L)^{ab}_{ij} = -(\Delta_R)^{ab}_{ij} = 
 \Delta_8 \left(\delta^a_i\delta^b_j
 - \delta^a_j\delta^b_i \right).
\ee
The inverse Nambu Gorkov propagator is not diagonal 
in color and flavor. It becomes diagonal in the space 
spanned by the 9 color-flavor matrices
\be
\label{basis}
 (v^A)^{ai} = \left(\begin{array}{cc}
  \frac{1}{\sqrt{2}}(\lambda^A)^{ai} & 0\\
  0 & \frac{1}{\sqrt{2}}(\lambda^A)^{ai}\end{array}\right),
\ee
where $\lambda^0=\sqrt{2/3}$ and $\lambda^A,\,(A=1,\ldots,8)$
are the Gell-Mann matrices. In this basis it is straightforward
to compute the inverse of (\ref{sinv}). We find
\be 
\label{S_NG}
 S^{AB} = \frac{\delta^{AB}}{p_0^2-\epsilon_p^2-\Delta_A^2}
 \left(\begin{array}{cc}
     p_0+\epsilon  & -\Delta^A \\
    -\Delta^A      & p_0-\epsilon_p \end{array}\right),
\ee
with $\Delta^A=2\Delta_8$ for $A=0$ and $\Delta^A=-{\rm sym}(A)
\Delta_8$ for $A=1,\ldots,8$. Here, ${\rm sym}(A)=1$ for the symmetric
Gell-Mann matrices $A=(1,3,4,6,8)$ and ${\rm sym}(A)=-1$ for
the anti-symmetric matrices $A=(2,5,7)$. 

 In order keep the dependence on $\Sigma$ we have to perform
the calculation using the anomalous self energy in the rotated
vacuum 
\bea
\label{del_rot}
 (\Delta_L)^{ab}_{ij} &=&
 \Delta_8 \left( X^a_i X^b_j - X^a_j X^b_i \right), \\
 -(\Delta_R)^{ab}_{ij} &=&
 \Delta_8 \left( Y^a_i Y^b_j - Y^a_j Y^b_i \right), 
\eea
with $X\in SU(3)_L$ and $Y\in SU(3)_R$. The Nambu-Gorkov 
propagator for left handed fermions is diagonal in a basis 
spanned by the color-flavor matrices
\be
\label{basis_rot}
 (\tilde{v}_L^A)^{ai} = \left(\begin{array}{cc}
  \frac{1}{\sqrt{2}}(\lambda^A X^T)^{ai} & 0\\
  0 & \frac{1}{\sqrt{2}}(\lambda^A X^\dagger)^{ai}\end{array}\right),
\ee
with a similar set of matrices $(\tilde{v}_R^A)^{ai}$ which
diagonalize the propagator for right handed fermions. In the 
basis (\ref{basis_rot}) the fermion propagator in the rotated 
CFL vacuum has exactly the same form (\ref{S_NG}) that it had
in the ordinary CFL vacuum (\ref{del_cfl}). The dependence
on $X,Y$ comes in when we calculate diagrams with external
color or flavor currents. In that case we have to take
matrix elements of the external current between the basis
states $(\tilde{v}_L)$ and $(\tilde{v}_R)$.

 We can now calculate a one-loop diagram with insertions
of $MM^\dagger$ and the gauge field $A_0$. We find
\be 
\Delta {\cal E} =  \frac{m_D^2}{2p_F} {\rm Tr}\left[
  X^\dagger MM^\dagger X A_0^T\right].
\ee
In the same way, we also calculate diagrams with insertions
of $M^\dagger M$ and $Q$. Collecting all these terms
we get 
\be
{\cal E} = \frac{m_D^2}{2}
 {\rm Tr} \left[ \left(X^\dagger\mu_e Q X
   + X^\dagger \frac{MM^\dagger}{2p_F} X + A_0^T\right)^2
 + \left(Y^\dagger\mu_e Q^\dagger Y
   + Y^\dagger \frac{M^\dagger M}{2p_F} Y + A_0^T\right)^2
\right] .
\ee
Similar to the calculation of $f_\pi$ it is essential here
to take into account the mixing with the gauge field. Without
the $A_0$ field we would conclude that there is no dependence
on the flavor matrices $X,Y$. We can now integrate out the 
gauge field $A_0$. We get
\bea 
\label{e_chir}
\Delta {\cal E} &=& \frac{m_D^2}{4} {\rm Tr}\left[
 \left(  \left(
 \mu_e Q +\frac{MM^\dagger}{2p_F}\right)\Sigma-
 \Sigma \left( \mu_e Q^\dagger +\frac{M^\dagger M}{2p_F}\right)\right)
 \right. \nonumber \\
 & & \hspace{3cm}\left.
 \left(  \left(
 \mu_e Q^\dagger +\frac{M^\dagger M}{2p_F}\right)\Sigma^\dagger-
 \Sigma^\dagger \left( \mu_e Q +\frac{MM^\dagger}{2p_F}\right)\right)
 \right],
\eea
where $\Sigma=XY^\dagger$. We note that after integrating out the 
gauge field the vacuum energy (\ref{e_chir}) only depends on the chiral 
field $\Sigma$ and not on $X$ and $Y$ separately. Using $f_\pi=m_D$ we
observe that (\ref{e_chir}) contains the terms required to complete 
the covariant derivative in (\ref{leff2}).

%%%%%%%%%%%%%%%%%%%%%%%%%%%%%%%%%%%%%%%%%%%%%%%%%%%%%%%%%%%%%%%%%%%%%%%%%

\newpage\noindent

\newpage

\newpage 

\newpage 

\end{document}